\documentclass[11pt]{article}   	
\usepackage{geometry}                		
\usepackage{graphicx}				
\usepackage{amssymb}
\usepackage{amsmath}
\usepackage{authblk}
\usepackage[table]{xcolor}
\usepackage{multirow}
\usepackage{graphicx}
\usepackage[normalem]{ulem}
\usepackage[english]{babel}
\usepackage{appendix}
\usepackage{float}
\usepackage{xcolor}

\newcommand{\be}{\begin{equation}}
\newcommand{\ee}{\end{equation}}

\usepackage[utf8]{inputenc}
\usepackage{xspace}
\usepackage{lineno}

\usepackage[hypertexnames,
    pdftex,%
    colorlinks,%
		citecolor=blue,%
    hyperindex,%
    plainpages=false,%
    bookmarksopen,%
    bookmarksnumbered%
  ]{hyperref}
\usepackage[numbers,sort&compress]{natbib}
\DeclareRobustCommand\GeV{\ensuremath{\mathrm{GeV}}\xspace}

\definecolor{eM}{hsb}{0.3,0.5,1}
\definecolor{eS}{hsb}{0,1,0.6}

\newcommand\sBD[1]{\cellcolor{eM}\color{eS}\textbf{#1}}
\newcommand\sB[1]{\color{eS}\textbf{#1}}

\graphicspath{{figures/}}

\usepackage{color}
\usepackage[normalem]{ulem}
\definecolor{pncolor}{rgb}{0,0.1,0.7}
\definecolor{ascolor}{rgb}{1,0,1}
\definecolor{nacolor}{rgb}{1,0,0}
\definecolor{wscolor}{rgb}{0,0.6,0.2}

\DeclareRobustCommand\pnout{\bgroup\markoverwith{\color{pncolor}{\rule[0.4ex]{2pt}{0.8pt}}}\ULon}
\DeclareRobustCommand\asout{\bgroup\markoverwith{\color{ascolor}{\rule[0.4ex]{2pt}{0.8pt}}}\ULon}
\DeclareRobustCommand\naout{\bgroup\markoverwith{\color{nacolor}{\rule[0.4ex]{2pt}{0.8pt}}}\ULon}
\DeclareRobustCommand\wsout{\bgroup\markoverwith{\color{wscolor}{\rule[0.4ex]{2pt}{0.8pt}}}\ULon}

\title{\bf Prospects for measurements of the longitudinal 
proton structure function {\boldmath $F_{L}$} at the Electron Ion Collider}
\author[1]{Javier Jim\'enez-L\'opez \thanks{javier.jimenezl@edu.uah.es (Corresponding author)}}
\author[2]{Paul R. Newman \thanks{paul.newman@cern.ch}}
\author[3]{Katarzyna Wichmann \thanks{katarzyna.wichmann@desy.de}}
\affil[1]{\small \it Universidad de Alcal\'a, Departamento de F\'isica y Matem\'aticas, Facultad de Ciencias, 28805 Alcal\'a de Henares, Madrid, Spain.}
\affil[2]{\small \it School of Physics and Astronomy, University of Birmingham, B15 2TT, UK}
\affil[3]{\small \it Deutsches Elektronen-Synchrotron DESY, Germany}
\date{DESY-24-218 \\
December 2024}
\begin{document}
\maketitle
\begin{abstract}
We explore the potential for extracting the longitudinal proton 
structure function $F_{L}$
at the future Electron-Ion Collider (EIC)
through a Rosenbluth separation method. 
The impacts of differing assumptions on sample sizes, systematic 
uncertainties and beam energy scenarios are investigated.
With a sufficiently large
number of centre of mass energy configurations and well-controlled
systematics,
the EIC will measure $F_{L}$ to an unprecedented precision,
even with relatively modest luminosities. The accessible
kinematic range complements both fixed target and
HERA data. In the most optimistic scenarios, the EIC data will be a
highly competitive direct probe of the proton gluon density.
\end{abstract}

\section{Introduction}
\label{sec:intro}

The investigation of the internal structure of the proton is 
an important topic in high-energy physics, offering deep 
insights into the composition and dynamics of 
strongly interacting matter. Central to these studies has been the measurement of cross sections for Deep Inelastic Scattering (DIS) 
and thus of the proton structure functions, which in turn
provide insight into the distributions of quarks and gluons 
in the nucleon \cite{10.1093/acprof:oso/9780198506713.001.0001}. The Hadron-Electron Ring Accelerator (HERA) was the first and
to date the only electron-proton colliding beam facility. It allowed measurements of inclusive DIS cross sections, and extraction of the structure functions $F_{2}$, $F_{L}$ and $x F^{\gamma Z}_3$\cite{H1:2015ubc}. Among these, the longitudinal structure 
function $F_L$ plays a special role through its direct, approximately
linear, dependence on the gluon density, which has been 
widely exploited
in fits for proton parton densities~\cite{Alekhin:2017kpj,Hou:2019efy,Bailey:2020ooq,NNPDF:2021njg} and phenomenological 
models~\cite{Cooper-Sarkar:1987cnv,Boroun:2012bje,Zijlstra:1992qd}.
Knowledge of both the $F_2$ and $F_L$ 
structure functions equivalently enables
extractions of the 
ratio $R$
between the inclusive cross sections for 
longitudinally to transversely polarised virtual photons to
interact with the proton,
according to
$R = F_L / (F_2 - F_L)$.

The extractions of $F_{L}$ at HERA \cite{H1:2013ktq,ZEUS:2014thn} are limited by statistics and cover a restricted range in the Bjorken variable $x$ and the squared four-momentum transfer $Q^2$.
It has also been possible to obtain 
$F_L$ for the proton at low $Q^2$ from numerous fixed-target
DIS experiments \cite{BENVENUTI1989485,ARNEODO19973,E140X:1995ims,Tvaskis:2006tv,Whitlow:1991uw,PhysRevC.97.045204,JeffersonLabHallCE94-110:2004nsn}.
%
The Electron-Ion Collider~\cite{Accardi:2012qut} (EIC), 
which is expected to begin science operations at Brookhaven 
National Laboratory in the early 2030s,
promises to revolutionise our understanding of the 
internal structure of hadrons. 
With its capability to operate at high luminosities and 
to cover a wide range of beam energy configurations, the EIC will provide information in regions of $x$ and $Q^2$ that are 
not well covered by previous experiments and are well suited to
$F_L$ extractions. 
In this study, we assess the EIC $F_{L}$ capabilities by
analysing simulated `pseudodata' with varying assumptions
on systematic precision and beam energy configurations. 



\section{EIC pseudo-data} \label{subsec:pseudo}

The accelerator and detector designs for the EIC are currently undergoing intense development, but the overall specification in terms of beam energy ranges and 
instrumentation coverage and performance
are already well-established~\cite{AbdulKhalek:2021gbh}. 
The studies presented in this paper are based on simulated data points or 'pseudodata'
which are derived from that baseline configuration. 
Our approach to producing inclusive DIS EIC pseudodata 
follows that of \cite{Armesto:2023hnw}, which in turn took 
binning schemes based on those in
the ATHENA detector proposal~\cite{ATHENA:2022hxb}. The ATHENA 
collaboration has since merged with ECCE~\cite{adkins2023design} to create the 
ePIC collaboration, which is now rapidly converging
towards a design for a first EIC detector. 
While some specifics of the detectors have evolved, the overall
expected kinematic range, kinematic variable resolutions and achievable experimental
precision are largely independent of the detailed detector design and this
study is thus applicable to any general purpose EIC experiment. 

\noindent As summarized in Table~\ref{tab:samples}, pseudo-data are produced with 
integrated luminosities corresponding to 
expectations for one year of  
data collection in each of the five expected beam 
configurations at the EIC~\cite{AbdulKhalek:2021gbh},
giving five different center-of-mass energies $\surd s$. 
For each beam configuration, pseudo-data are produced at five logarithmically spaced $x$ values per decade over the 
inelasticity range \(0.005 < y < 0.96\), 
matching the expected experimental resolutions~\cite{ATHENA:2022hxb}. 
The EIC resolution in \(Q^2\) is expected to be significantly better than that
of the HERA experiments, so for ease of comparison, we adopt the   
\(Q^2\) values from the measurements of \(F_L\) in~\cite{H1:2013ktq, ZEUS:2014thn}. 

The central values of the pseudodata cross sections are initiated using 
NNLO theoretical predictions based on the HERAPDF~\(2.0\) parton densities~\cite{H1:2015ubc}, with the value for each data point 
randomly smeared using samples from Gaussian distributions that reflect the assumed 
experimental uncertainties. 
Two different scenarios are considered. The first, referred to here as the `conservative
scenario' is based largely on the systematic precision achieved at HERA,
and follows the considerations in~\cite{AbdulKhalek:2021gbh}, 
as also adopted in the studies by  
ATHENA collaboration, and subsequently used to study EIC collinear PDF sensitivities~\cite{Armesto:2023hnw} and \(\alpha_s\) measurements~\cite{Cerci:2023uhu}. 
In the moderate $y$ range studied here, 
the data points have a point-to-point uncorrelated systematic uncertainty 
of 1.9\(\%\).
A normalisation uncertainty
of 3.4\(\%\) is also included for each data set, which is not correlated between 
different beam energy configurations, such that the total uncertainty 
attributed to each data point is 3.9 \(\%\).
More recent estimates of the achievable EIC precision suggest that a much better performance 
will be obtained than that in the conservative scenario. There is also expected to be
some correlation between the normalisation uncertainties at different beam energies. 
We therefore also consider a second `optimistic' scenario, which follows the assumption 
in \cite{PhysRevD.105.074006} that the total uncertainty that
is uncorrelated between data points at different beam energies is 1\(\%\). 
At the time of writing, the  
rather different assumptions in the conservative and optimistic scenarios can
be seen as extreme cases with the final achieved precision likely to lie between them. 


\begin{table}[htb!]
\centerline{
\begin{tabular}{|c|c|c|c|}
\hline
$e$-beam energy (GeV) & $p$-beam energy (GeV) & $\sqrt{s}$ (GeV) & Integrated lumi (fb$^{-1}$) \\ \hline 
18 & 275 & 141 & 15.4 \\
10 & 275 & 105 & 100 \\
10 & 100 &  63 & 79.0 \\
5 &  100 &  45 & 61.0 \\
5 &   41 &  29 &  4.4 \\ \hline
\end{tabular}}
\vskip 0.4cm
\caption{Beam energies, centre-of-mass energies and 
integrated luminosities assumed for the 
different EIC configurations considered.
}
\label{tab:samples}
\end{table}


\section{Extraction of $F_{L}$} \label{subsec:fit}
\label{sec:extraction}

The neutral current (NC) $e^{\pm}p$ DIS cross sections are given by a linear combination
of generalised structure functions~\cite{10.1093/acprof:oso/9780198506713.001.0001}.
 For $Q^2 << M_{Z}^2$ (mass of the Z boson squared), the inclusive cross section for NC DIS can be written in terms of the two structure functions $F_{L}$ and $F_{2}$ as

\begin{equation} \label{cross_sec}
    \dfrac{{\rm d}^2 \sigma^{e^{\pm}p}}{{\rm d}x {\rm d}Q^2} = \dfrac{2 \pi \alpha^2 Y_{+}}{xQ^4} \left[ F_{2}(x, Q^2) - \dfrac{y^2}{Y_{+}} F_{L}(x, Q^2) \right] = \dfrac{2 \pi \alpha^2 Y_{+}}{xQ^4} \sigma_{r}(x, Q^2, y) \ ,
\end{equation}

where $\alpha$ is the fine structure constant, 
$Y_{+} = 1 + (1 - y)^2$, and $\sigma_{r}$ is usually referred to as the reduced cross section. 

Eq.~\ref{cross_sec} implies that there is a linear relationship between the reduced cross section  and $y^2 / Y_{+}$, which is a function only of the inelasticity of the process, adjustable 
at fixed $x$ and $Q^2$
by changing the centre of mass energy and exploiting the relationship
$Q^2 \simeq s x y$.
Using measurements of $\sigma_{r}$ at different $s$,  the values of $F_{2}$ and $F_{L}$ can thus be separately obtained in a model independent way 
as the free parameters of a linear fit.
This Rosenbluth-type separation technique \cite{PhysRev.79.615}
has been employed in fixed target data 
and at HERA
and has also recently been applied to the extraction of the diffractive longitudinal structure function at the EIC~\cite{PhysRevD.105.074006}.
For each $(x, Q^2)$, we thus apply a fit of the form
\begin{equation}\label{fit}
    \sigma_r(x,Q^2,y) = F_2(x,Q^2) - \frac{y^2}{Y_+} F_L(x, Q^2) \ ,
\end{equation}
where $F_2(x,Q^2)$ and $F_L(x,Q^2)$ are free parameters, obtained
through a $\chi^2$ minimisation
implemented in PYTHON using SciPy \cite{Virtanen:2019joe}.
Equation~\ref{cross_sec} can also be fitted using a Bayesian method to obtain $F_L$, as done 
by the ZEUS Collaboration~\cite{ZEUS:2014thn}. 
We used the BAT package~\cite{Schulz:2021BAT} to cross-check the results 
using this method, obtaining 
perfect agreement.



\begin{figure}[htb]
    \centering  
     \includegraphics[scale = 0.28]{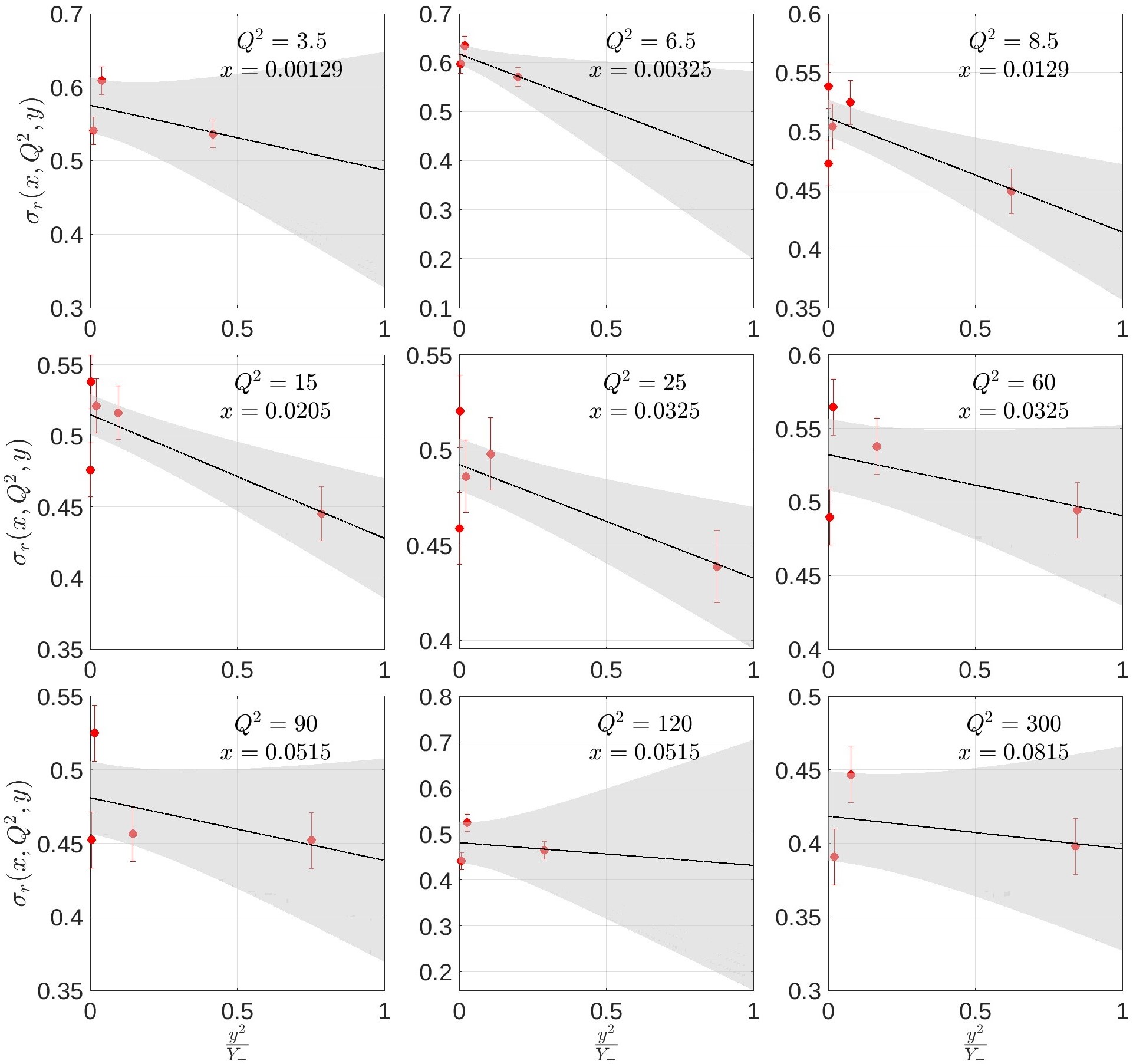} 
    \caption{Results of the fitting procedure for 9 example 
    $x$ and $Q^2$ values in the conservative scenario. The simulated reduced cross section measurements are shown in red, where each point corresponds to a different center-of-mass energy and the vertical error bars represent the total uncertainties described in Section~\ref{subsec:fit}. The $68 \; \%$ confidence bands for each fit are depicted in light grey. Note that there is a suppressed zero on the vertical axis. 
     }
    \label{fits}
\end{figure}

Figure~\ref{fits} depicts fits in the 
conservative scenario at $9$ example values of $x$ and $Q^2$, 
together with the $68 \; \%$ confidence bands for each fit.
Due to the details of the binning scheme adopted here, 
the range in $y^2 / Y_+$ varies substantially between bins.
Different combinations of $Q^2$ and $x$ lead to data points 
within the acceptable $y$ range at 
$3$, $4$, or $5$ different $\surd s$ values.
For $Q^2$ and $x$ combinations with less than three such data points, no 
extraction of $F_{L}$ is attempted. 


The pseudo-data smearing procedure introduces randomness into the 
fit inputs, which is reflected in the outputs as 
fluctuations in both the
extracted $F_L$ values and their uncertainties, which are obtained from 
variations of the parameter resulting in a $\Delta \chi^2 = 1$. 
The $F_L$ uncertainty results are therefore samples from distributions of
possible outcomes which are often quite broad.
In order to sample the distribution of possible uncertainty outcomes
in a more systematic way, we adopt the method 
introduced in~\cite{PhysRevD.105.074006} and 
further described in Section~\ref{sec:ave},
whereby the results are averaged over multiple replicas of the procedure.
With sufficiently large numbers of replicas
(1000 is used by default), both the mean 
and the variance 
tend towards results that can be considered as 
expectation values for the simulated scenario. 

In the following sections, we only consider 
bins where the absolute uncertainty of $F_L$ averaged over 
the 1000 replicas is smaller than 0.3. 
This choice results in more data points surviving  
in the optimistic than the conservative scenario
since the overall level of the uncertainties is smaller.
Particularly for small $Q^2$, it also tends to exclude
some data points at high $x$ with large 
uncertainties that might be recovered in a 
future analysis.
\section{Results}
\label{sec:results}

\subsection{Example $F_{L}$ replicas}
\label{sec:replicas}

\begin{figure}[htbp]
    \centering  
    \includegraphics[scale = 0.35]{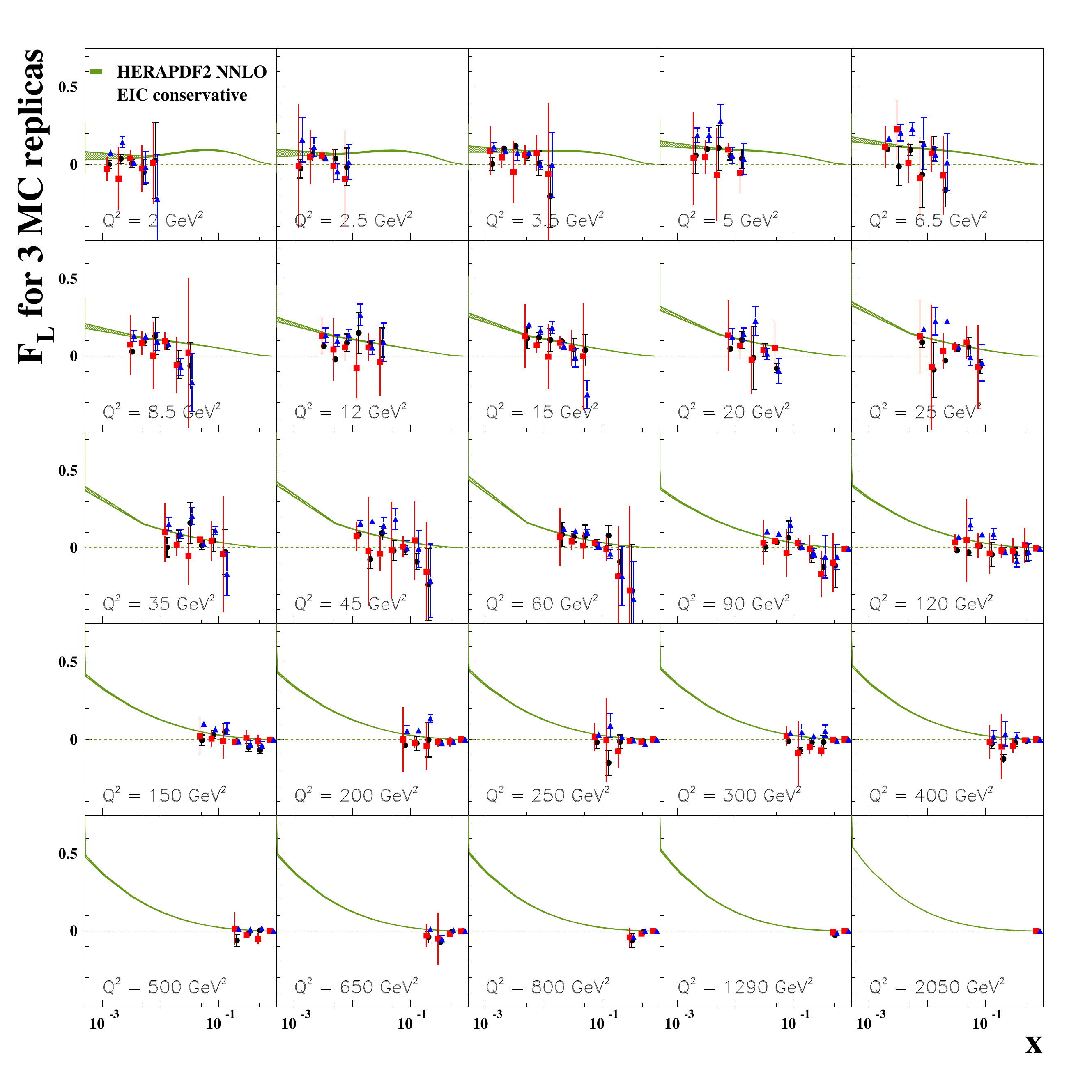}
    \caption{Simulated extractions of the longitudinal structure function $F_{L}$ for three different 
    pseudo-data replicas in the conservative scenario, with different colours
    corresponding to different replicas, whose $x$ positions are slightly
    shifted for visibility. The theoretical predictions from HERAPDF$2.0$ NNLO are also shown as bands whose widths correspond to the uncertainties on
    the predictions.}
    \label{5_repP2}
\end{figure}

\begin{figure}[htbp]
    \centering  
    \includegraphics[scale = 0.35]{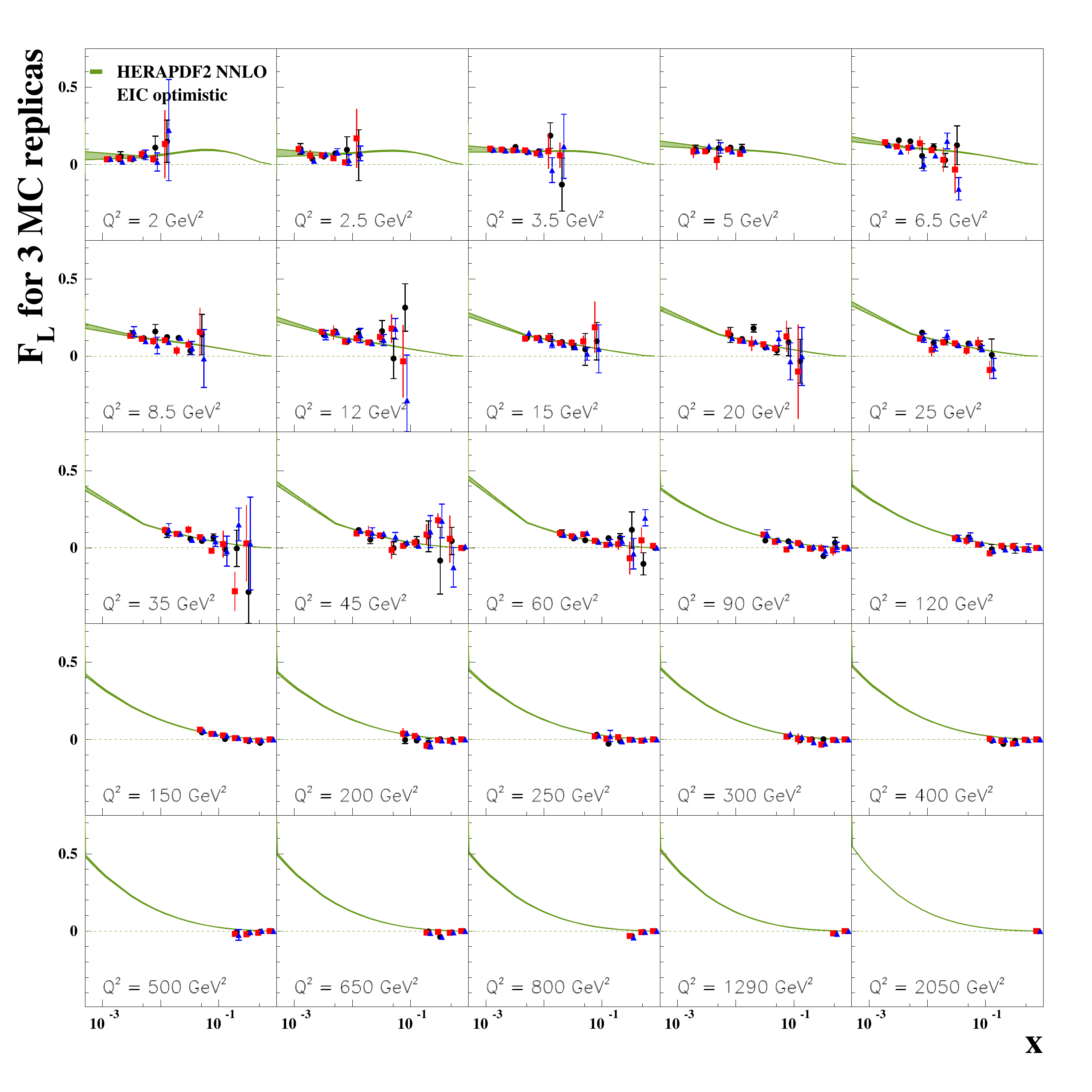}
    \caption{Simulated extractions of the longitudinal structure function $F_{L}$ for three different 
    pseudo-data replicas in the optimistic scenario, with different colours
    corresponding to different replicas, whose $x$ positions are slightly
    shifted for visibility. The theoretical predictions from HERAPDF$2.0$ NNLO are also shown as bands whose widths correspond to the uncertainties on
    the predictions.}
    \label{5_repO2}
\end{figure}

To sample the distribution of possible outcomes for the expected 
values and uncertainties on $F_{L}$ in a statistically meaningful way, 
we performed the smearing and fitting procedure described in 
sections~\ref{subsec:pseudo} and~\ref{sec:extraction} multiple times, leading to a set of
replica results, each of which is an independently created pseudodata set.
Figures ~\ref{5_repP2} and ~\ref{5_repO2}  show the resulting
longitudinal structure function $F_{L}$ 
overlaid for three randomly chosen example replicas, 
for the conservative and optimistic scenario, respectively.

The spread between different replicas, arising from the random smearing
of the reduced cross section data to simulate their uncertainties, propagates
into the $F_L$ pseudo-data, leading to substantially more precise results and a 
smaller spread between replicas in the optimistic than in the conservative scenario.
The theoretical predictions calculated with HERAPDF2.0 NNLO PDF set are also shown
in the figures,
and describe the pseudo-data well by construction.  

\subsection{Averaging over $F_{L}$ replicas}\label{sec:ave}

\begin{figure}[htbp]
    \centering  
    \includegraphics[scale = 0.375]{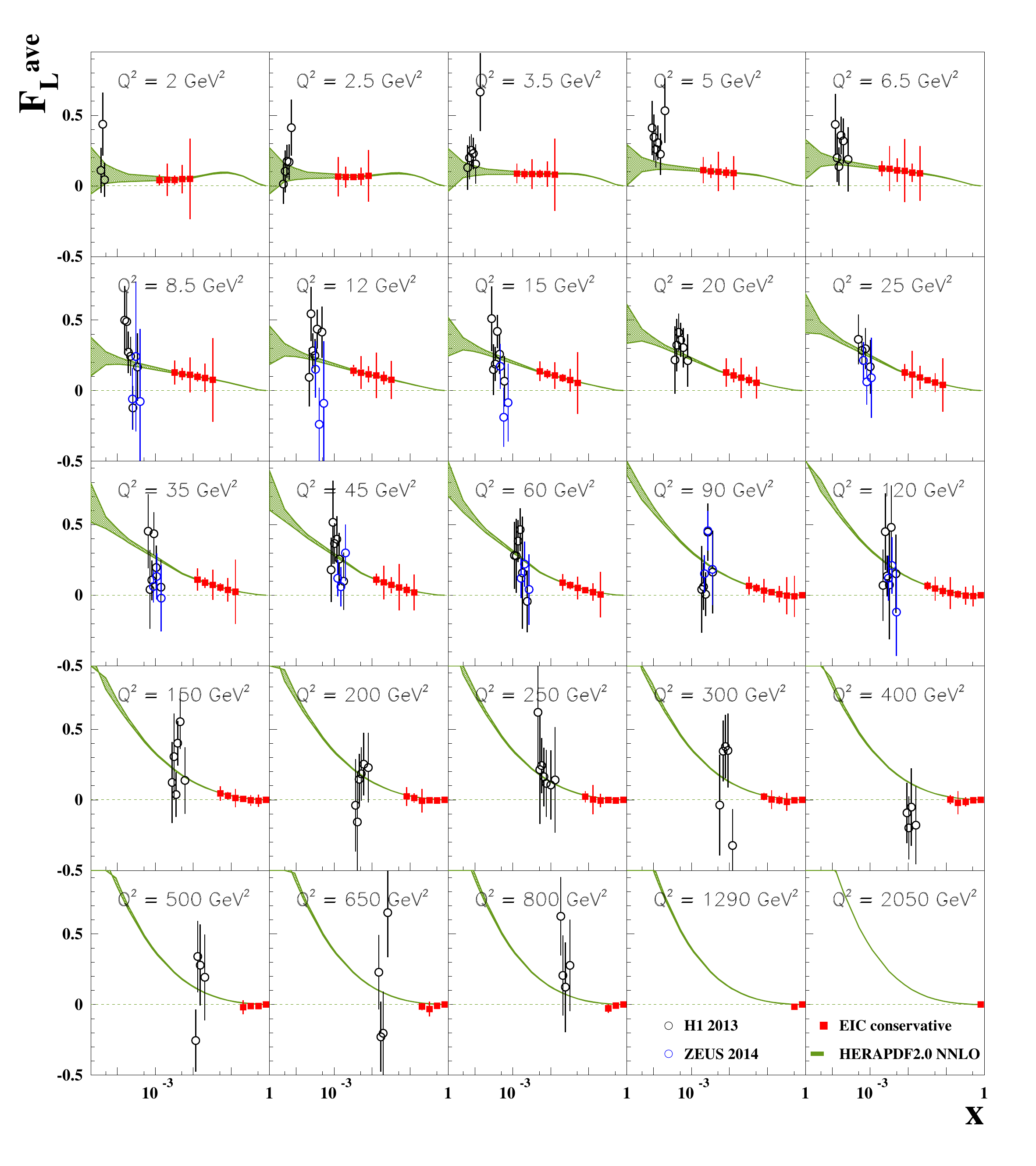}
    \caption{Simulated extractions of the longitudinal structure function $F_{L}$ averaged over $1000$ replicas for the conservative EIC uncertainty scenario, shown together with data from the H1 and ZEUS collaborations, alongside the theoretical predictions from HERAPDF $2.0$ NNLO. The error bars on the points represent
    the total experimental uncertainties, while the width of the bands
    correspond to the uncertainties on the theoretical predictions.}
    \label{AvgP}
\end{figure}

\begin{figure}[htbp]
    \centering  
    \includegraphics[scale = 0.375]{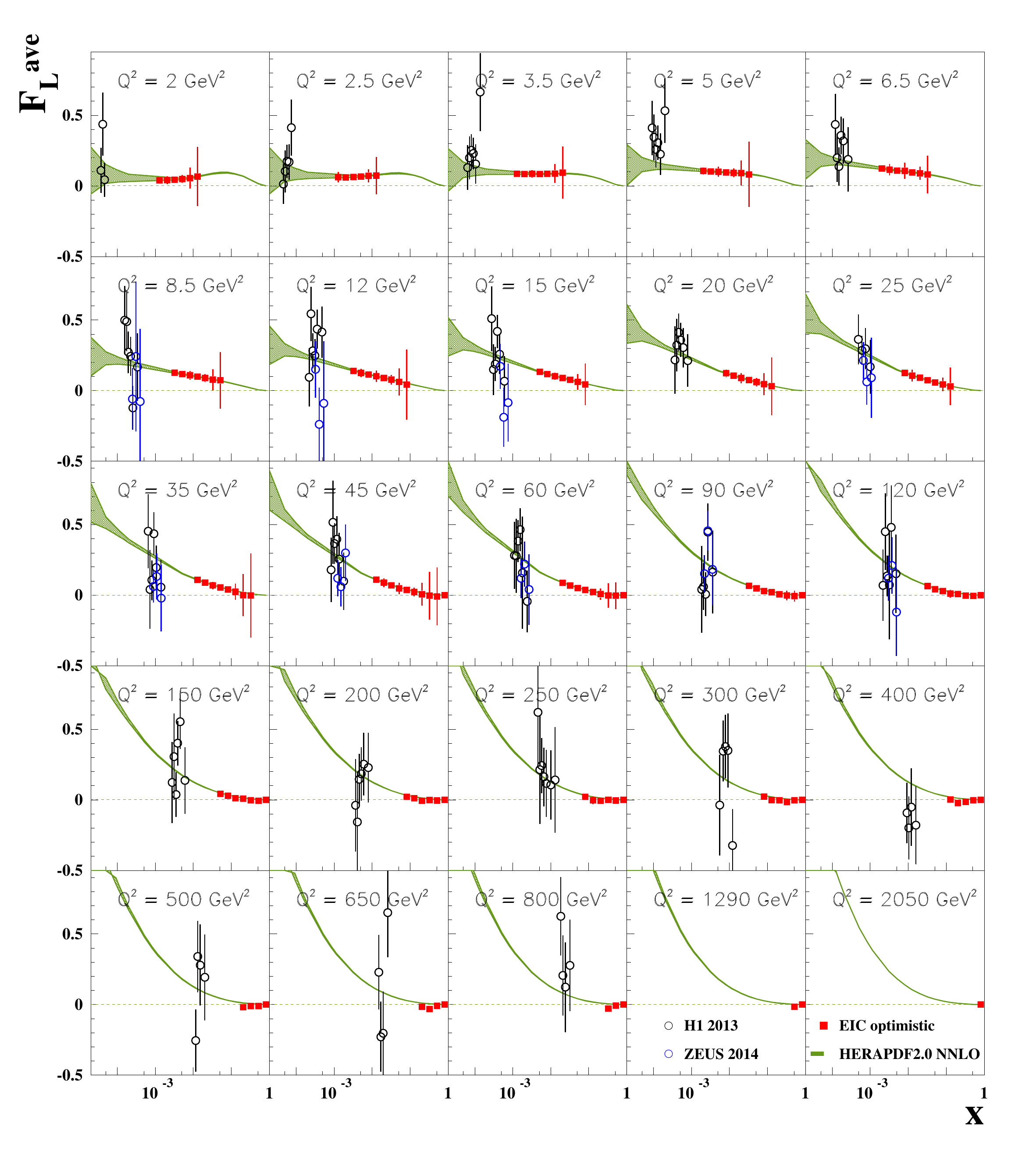}
    \caption{Simulated extractions of the longitudinal structure function $F_{L}$ averaged over $1000$ replicas for the optimistic EIC uncertainty scenario, shown together with data from the H1 and ZEUS collaborations, alongside the theoretical predictions from HERAPDF $2.0$ NNLO. The error bars on the points represent
    the total experimental uncertainties, while the width of the bands
    correspond to the uncertainties on the theoretical predictions.}
    \label{AvgO}
\end{figure}

To obtain a robust expectation for the predicted uncertainties on \( F_{L} \), we used the averaging procedure described in \cite{PhysRevD.105.074006},
which yields a single pseudodata point for \( F_{L} \) at each $x$ and $Q^2$ value, averaged over 1000 replicas. 
The average and variance were calculated as:

\begin{equation}\label{avg_proc_1}
    \overline{v} = \frac{S_{1}}{N}, \quad (\Delta v)^2 = \frac{S_{2} - S^2_{1}/N}{N - 1},
\end{equation} 
where \( S_{n} = \sum_{i = 1}^{N} v^n_{i} \) and \( v_{i} \) is the value of \( F_{L} \) in the \( i \)-th MC sample. 
The results are shown in Figures~\ref{AvgP} and \ref{AvgO} for the conservative and optimistic scenarios, respectively. 
The fluctuations in the uncertainties between neighbouring data points
reflect the variations in the numbers of bins and their $y^2 / Y_+$
ranges included in the Rosenbluth separation fits, as illustrated
in figure~\ref{fits}. A smoother response could be achieved with a
modified binning scheme for the input pseudo-data at each $\surd s$,
for example a switch from an $(x, Q^2)$ to a $(y, Q^2)$ grid. 

The uncertainties on the averaged points, which can now be taken as a measure of the 
expected precision, can be compared with the uncertainties on the  
theoretical predictions, which 
are indicated through the widths of the corresponding bands
and are 
driven primarily by the uncertainty on the gluon density in the 
HERAPDF $2.0$ PDF set. 
In the optimistic
scenario, the precision of the pseudo-data is comparable to that of the 
theoretical predictions at low and intermediate $Q^2$, suggesting that such a measurement would provide a direct constraint that significantly 
improves our knowledge of
the gluon density in the proton. Although this is not 
so evidently the case in the conservative
scenario, the data would still be a useful ingredient in fits, as well as providing
a test of the consistency of the overall PDF framework. 
 
Measurements by the H1~\cite{H1:2013ktq} and ZEUS~\cite{ZEUS:2014thn} 
collaborations are also 
shown\footnote{Some of the ZEUS data points have slightly different $Q^2$ values 
from the H1 measurements, which are adopted here. 
The ZEUS data points have therefore been adjusted to the H1 
$Q^2$ values using factors obtained from 
HERAPDF2.0 NNLO. These factors were small, not exceeding 4\%.}
alongside the simulated EIC data in Figures~\ref{AvgP} and \ref{AvgO}.
The EIC data sample a region of $x$ between one and two orders of
magnitude higher than the HERA measurements, and also extend to larger $Q^2$,
ultimately reaching regions where the predicted $F_L$ values become very small. 
In both the optimistic and the conservative scenarios, the expected
precision of the EIC pseudodata is very much better than that of the HERA data,
a consequence of the much larger assumed integrated luminosities
and the larger number of $\surd s$ values in the Rosenbluth
fits. 
These figures illustrate that 
in both uncertainty scenarios, the EIC will measure the longitudinal 
structure function 
in a kinematic region that is complementary to that accessed at HERA and 
with a much improved precision. 

\begin{figure}[htb]
    \centering
    A) \includegraphics[scale = 0.18]{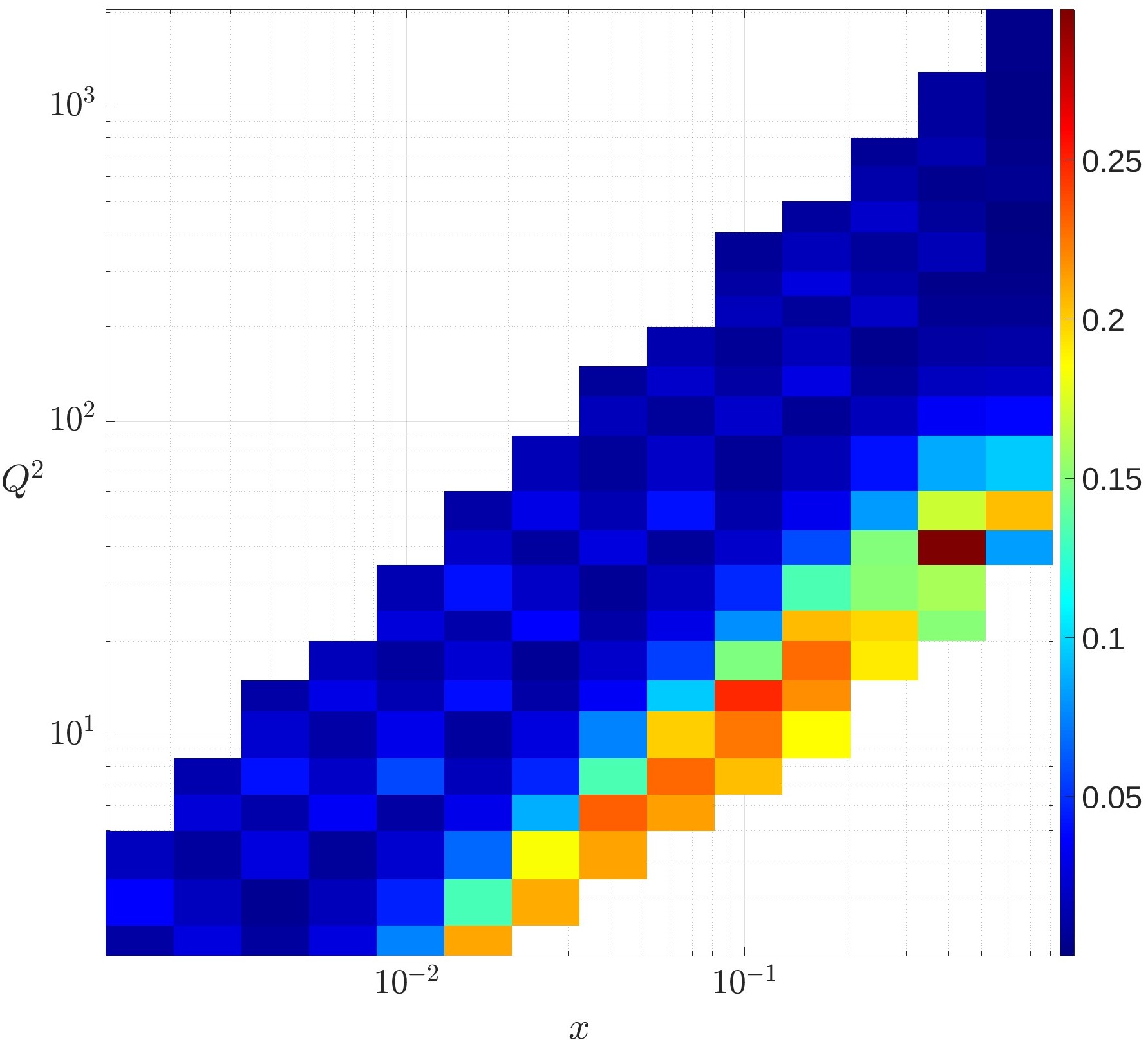}
    B) \includegraphics[scale = 0.18]{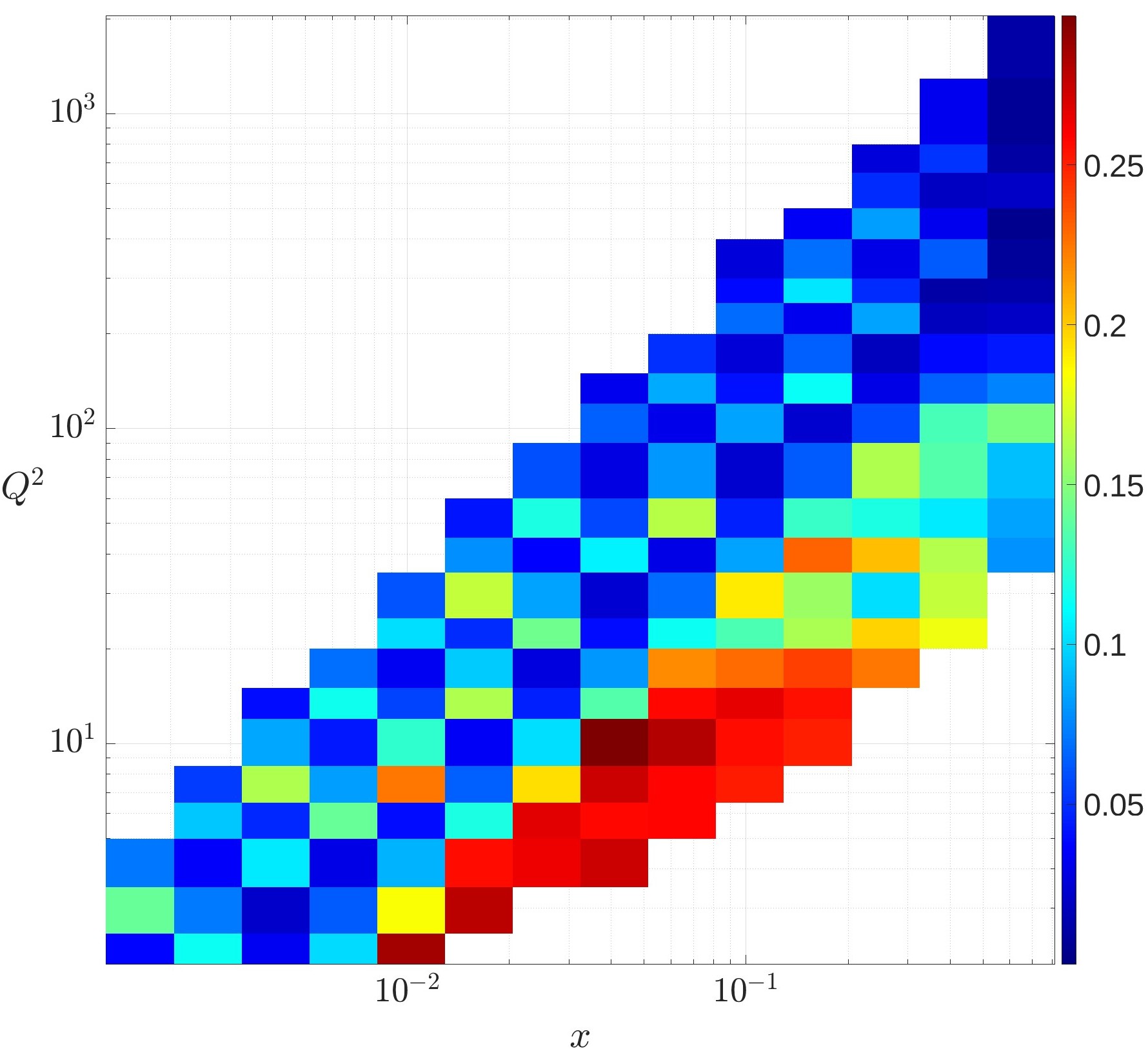}
    \caption{Uncertainties on the simulated EIC $F_L$ 
    measurements averaged
    over 1000 replicas.
    A) corresponds to the optimistic scenario  and B) to the conservative scenario, with the colours indicating the absolute uncertainties
    with the same scale in both cases. Points in $x$ and $Q^2$
    for which the 
    absolute uncertainty on $F_L$ is larger than 0.3 are not shown.}
    \label{Heat}
\end{figure}

In Figure~\ref{Heat} the expected precision on each 
$F_L$ data point is shown in the conservative and optimistic scenarios after carrying out the averaging procedure in 
Eq.~\ref{avg_proc_1}. 
The overall size of the uncertainties scales approximately linearly
with the systematic uncertainty assumptions in the two scenarios. 
Absolute values of the uncertainties at the level of 0.05 are obtained
across a wide kinematic range in the optimistic scenario, corresponding
to around 20\% precision where $F_L$ itself is relatively large. 
As can be seen in Fig~\ref{AvgP},
the HERA measurements have much larger uncertainties, in excess of 100\%
for over half of the data points. 
At the largest $x$ values, the quality of the simulated EIC measurements
deteriorates, as the lever-arm in $y^2 / Y_+$ gets shorter.
Bin-to-bin fluctuations are visible in the uncertainties, 
following similar
patterns between the optimistic and conservative scenarios, 
once again corresponding to the 
varying conditions in terms of numbers of data points and their $y^2/Y_+$
ranges, as illustrated in Figure~\ref{fits}. 




\subsection{Averaging $F_{L}$ over $x$}


The HERA $F_L$ measurements are often presented 
as a function of $Q^2$ after averaging over $x$. 
To compare the simulated EIC data in this projection, 
the $F_{L}$ pseudodata averaged over the $1000$  
replicas were further averaged over all $x$ values at each $Q^2$
using 
a simple weighted-mean procedure where the weights are 
derived from the uncertainties in each $F_{L}$ measurement.
The data are attributed to an averaged value of $x$, obtained
using the same weights as for the $F_{L}$ averaging.

\begin{figure}[htbp]
    \centering  
    A) \includegraphics[scale = 0.256]{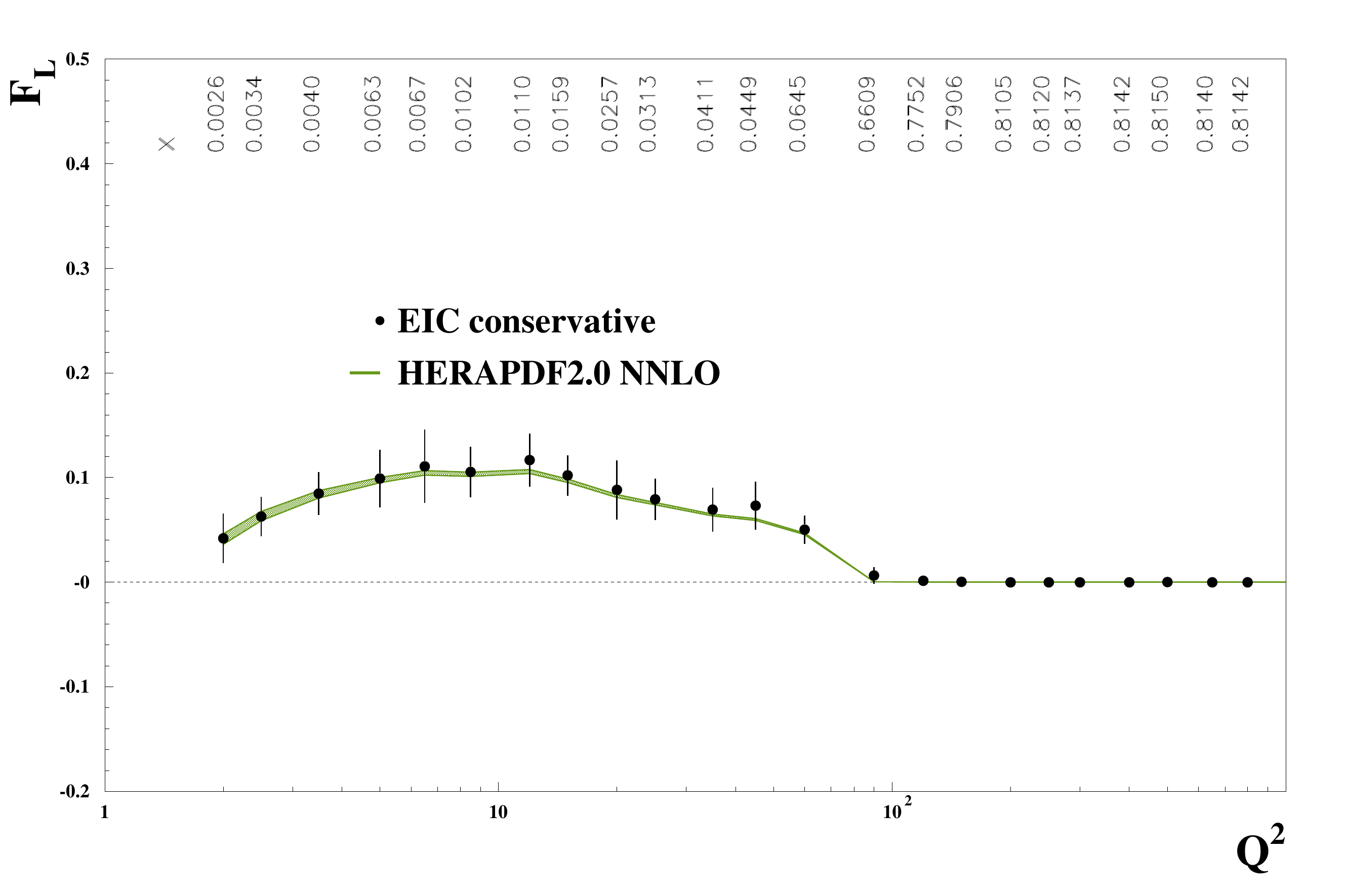}\\
    B) \includegraphics[scale = 0.6]{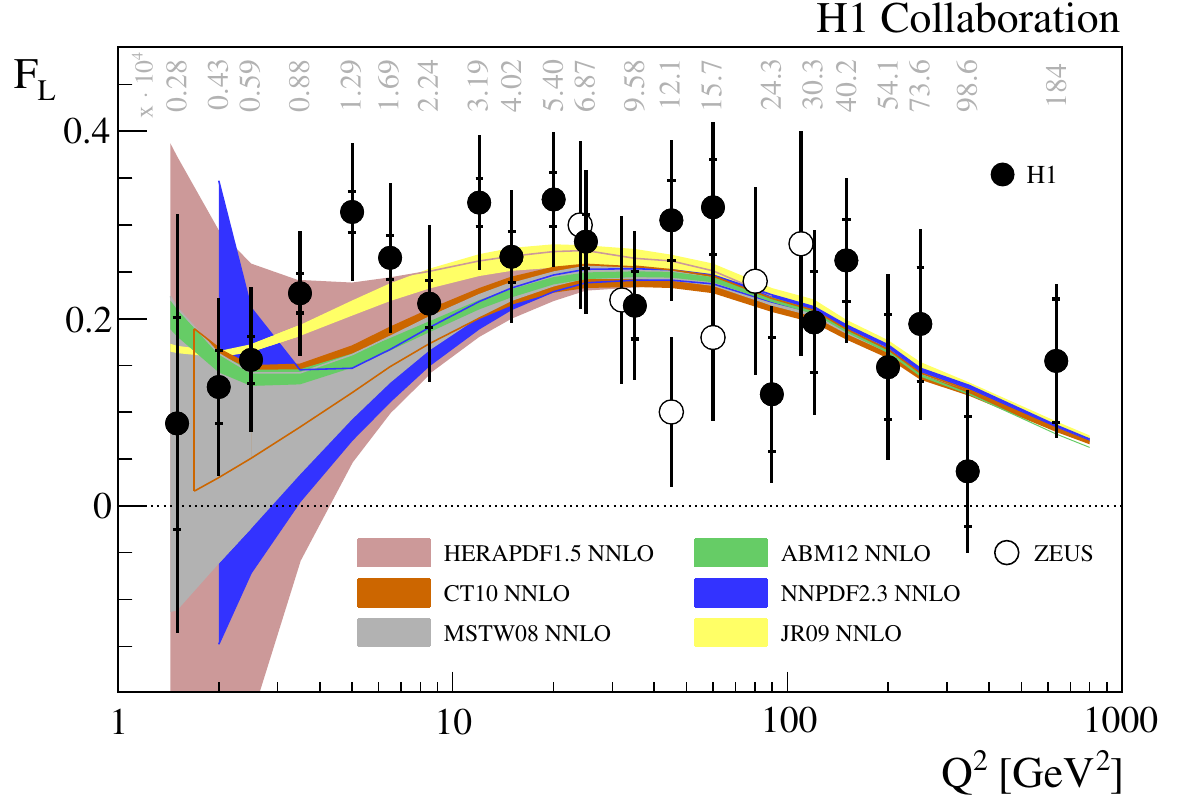}
    \caption{A) Simulated EIC measurements of the longitudinal structure function $F_{L}$ in the conservative scenario, averaged over 1000
    replicas, shown as a function of $Q^2$ and compared with the theoretical prediction given by HERAPDF$2.0$ NNLO.  B) Summary plot for 
    existing HERA measurements of $F_{L}$ averaged over $x$ and compared 
    with predictions at NNLO from various PDF sets \cite{H1:2013ktq}.
    The horizontal and vertical scales in A) and B) are chosen to be 
    identical for ease of comparison and in both cases the weighted-average
    $x$ value is indicated for each data point.}
    \label{Final}
\end{figure}

The results for $F_L(Q^2)$ in the conservative scenario are shown in Fig~\ref{Final} A), 
together with the average values of $x$ and the theoretical predictions based on HERAPDF2.0 NNLO.
The agreement of the pseudo-data and the predictions is very good,
as expected by construction.
The data are presented alongside an equivalent 
figure showing the HERA data \cite{H1:2013ktq}, 
with the 
axis scales chosen to be equivalent to
allow a direct comparison between the $Q^2$ ranges and the 
level of precision. 
Even for the conservative scenario shown here, 
the uncertainties on 
the EIC measurements are 
significantly smaller than those on the HERA data. 
The magnitude of $F_{L}$ for the EIC data points is
always smaller than that from HERA at the 
same $Q^2$ value, due to the dependence of $F_L$ on $x$
and the larger $x$ values sampled at EIC compared with HERA. 
It is clear that the EIC will be able to measure the longitudinal structure function $F_{L}$ 
with unprecedented precision and in so far unexplored kinematic regions.

\section{Alternative scenarios}
\label{sec:better}

Since the 
running schedule over time for the EIC is still far from certain,
we have studied the influence of different aspects, which may be 
helpful as input to the decision-making process. In addition to the 
influence of the systematic uncertainties, as studied 
in Section~\ref{sec:results}
through the 
difference between the optimistic and conservative scenarios,
it is
also interesting to investigate the influence 
of statistical uncertainties by varying the integrated luminosity
assumed for each of the different beam energy and $\surd s$ configurations. 
In place of the annual luminosities for each case assumed by default
(Table~\ref{tab:samples}), we have therefore repeated the full study
with only 1 fb$^{-1}$ for each of the beam energies. Although only a small
fraction of the peak luminosities expected for EIC, this is larger than the
total sample sizes obtained at HERA and may be a realistic EIC target
for early running, or for a dedicated running period with frequent beam
energy changes targeting physics studies such as that presented here
for $F_L$.


\begin{figure}[htb]
    \centering  
    \includegraphics[scale = 0.3]{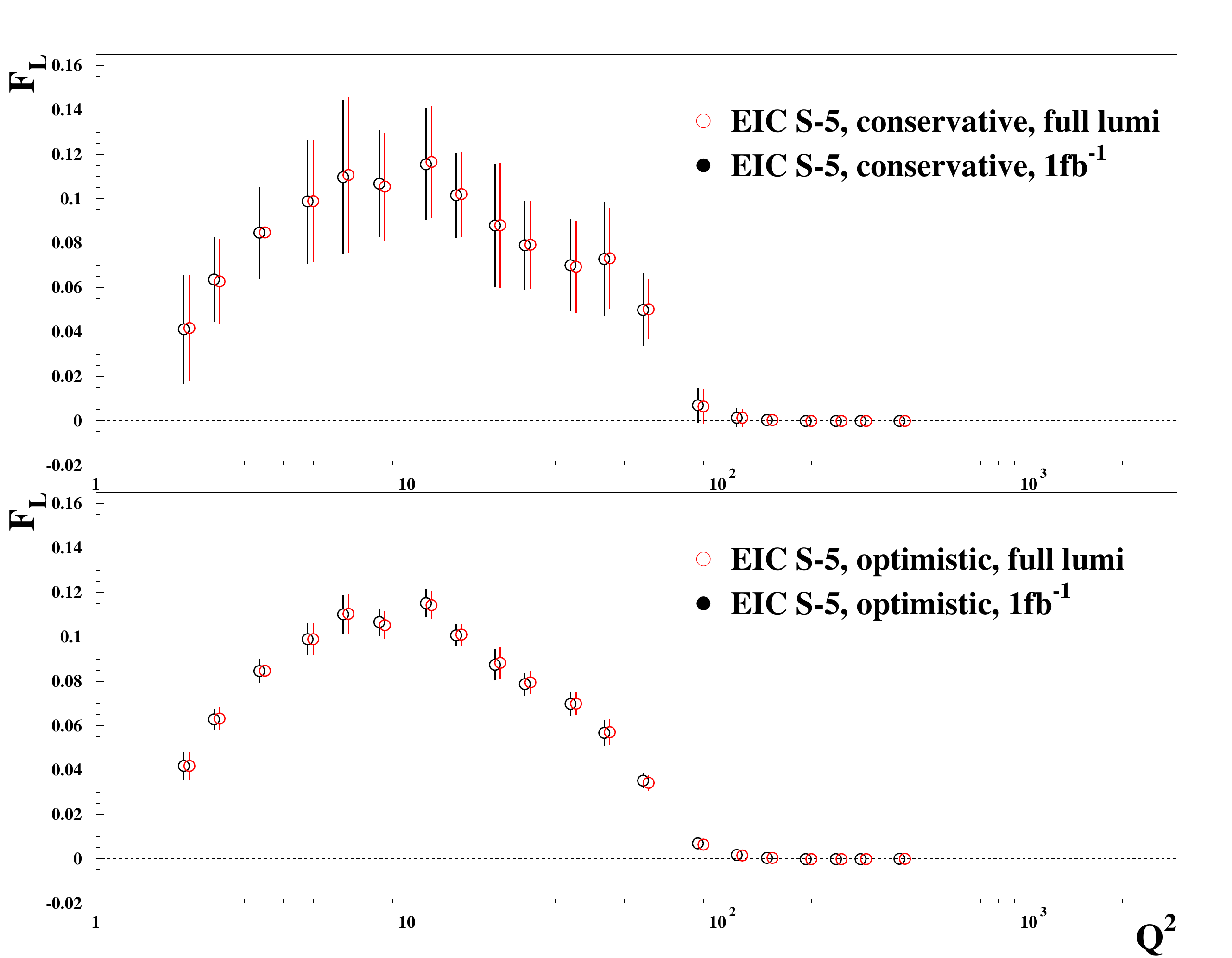}
    \caption{Simulated $F_{L}(Q^2)$ data based on 
    five beam energy configurations in the conservative (top) and 
    optimistic (bottom) scenarios. In both cases, comparisons are made
    between results assuming a full nominal year of integrated luminosity 
    in each configuration (Table~\ref{tab:samples}) and a scenario
    in which 1 fb$^{-1}$ is assumed for each configuration. The data
    points with the different luminosity assumptions are slightly
    off-set from one another for visibility.}
    \label{fl-1fb}
\end{figure}

The expected precision on the $F_L$ extraction as a function of $Q^2$
is compared for the different systematic and statistical uncertainty 
scenarios in figure~\ref{fl-1fb}. As for the two-dimensional projections,
there is a difference of a factor of around 3-4 between the results with
the conservative and optimistic scenarios, with the assumed uncertainties
propagating approximately linearly. The influence of
reducing the integrated luminosity is minor by comparison, confirming that
1 fb$^{-1}$ at each $\surd s$ is more than sufficient to achieve 
the required level of statistical precision and ensure that the $F_L$
extraction is systematically limited, even in the optimistic scenario 
with 1\% 
systematics.

\renewcommand{\arraystretch}{1.3}
\begin{table}[htb]
\begin{center}
\begin{tabular}{cr|*6r}
	& & \multicolumn{6}{c}{\(E_p\, [\GeV]\)} \\
 & & 41 & 100 & 120 & 165 & 180 & 275 \\
 \hline
 \multirow{3}{*}{\rotatebox{90}{\(E_e\, [\GeV]\;\)}}
 & \vbox to 12pt{}
 5 & \sBD{29} & \sBD{45} & 49       & \sB{57}  & 60       & 74 \\
 & 10 & 40 & \sBD{63} & 69 & \sB{81} & (85) & \sBD{105} \\
& 18 & 54 & \sB{85} & 93 & \sB{109} & 114 & \sBD{141} \\
\end{tabular}
\end{center}
\caption{Center-of-mass energies (in GeV) for various combinations of 
electron and proton beam energies. The default case studied here (S-5) with five 
configurations is illustrated with the green boxes. Additional 
combinations introduced in the S-9 scenario are depicted in red.
For the S-17 scenario, 
the full set of combinations is included except for the degenerate case
at $\sqrt{s}=85 \ {\rm GeV}$ (in parentheses).}
\label{tbl:ecm}
\end{table}

 We have additionally 
investigated the potential to improve the precision on $F_L$ by
increasing the 
number of beam energy configurations and hence $\surd s$ 
vales in the Rosenbluth decompositions. 
Following \cite{Armesto:2023hnw},
in addition to the default scenario with five configurations
(referred to in the following as S-5), we have also considered
different combinations of the same sets of electron and proton 
beam energies that lead to different $\surd s$ values. 
As illustrated in Table \ref{tbl:ecm}, we consider scenarios with
nine (S-9) or 17 (S-17) different configurations. Not all of the
combinations considered are necessarily technically realisable at the
EIC, so the choices made here should be considered only as indicative,
chosen in order to explore the potential improvements that might be
achievable. 


\begin{figure}[htb]
    \centering  
    \includegraphics[scale = 0.2]{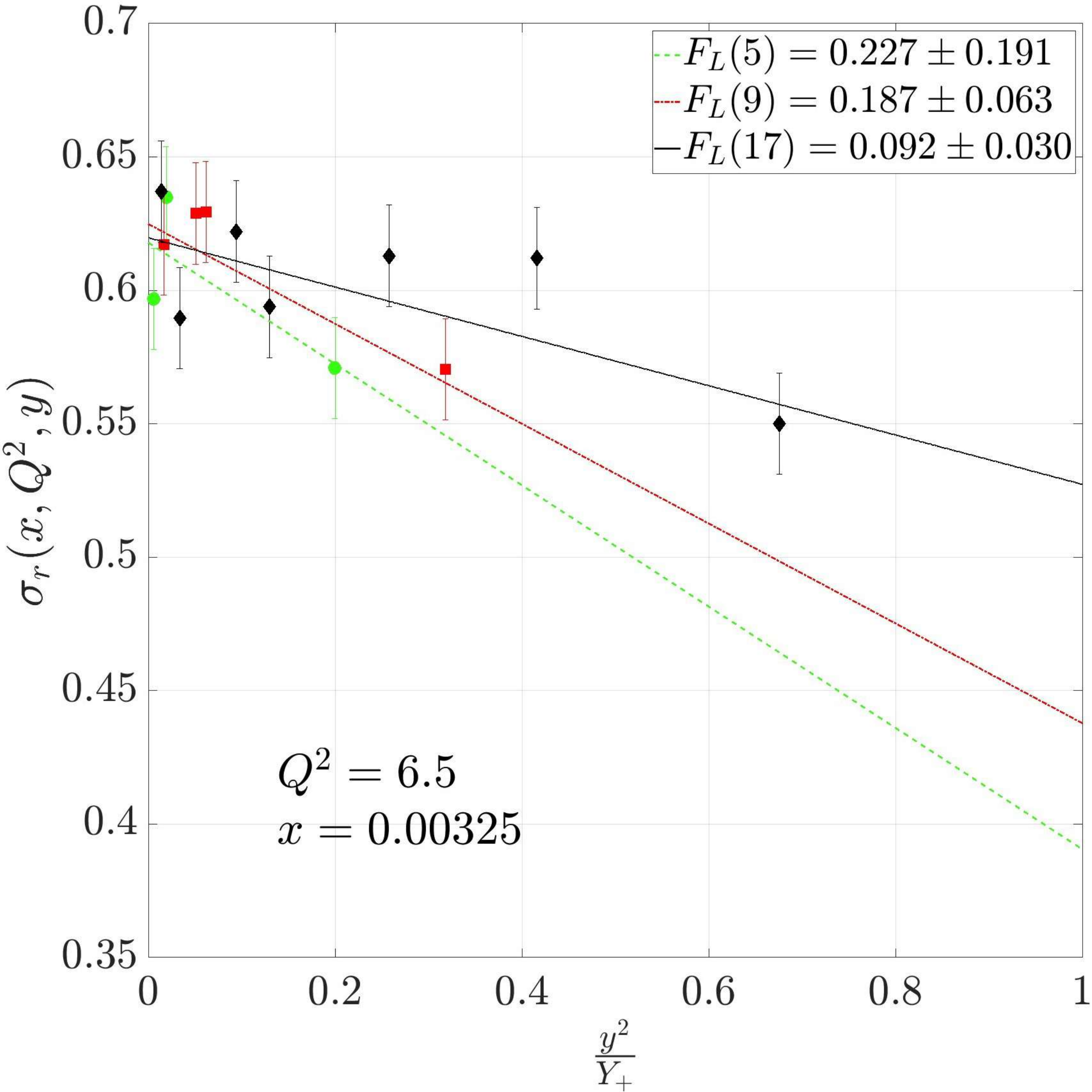} 
    \caption{Comparison of Rosenbluth decomposition fits to simulated
    EIC data in an example bin 
    defined by $Q^2 = 6.5 \; \text{GeV}^2$ and $x = 0.00325$
    under three different assumptions
    as to the beam energy configurations that are included. The points
    shown in green are available in the S-5 scenario, 
    and correspond to the fit shown as the green line and the 
    result indicated as `$F_L(5)$'. Additional points that become available
    in the S-9 scenario are shown in red, with the corresponding fit 
    to the green and red points shown in red and 
    indicated as `$F_L(9)$'. The black points
    are introduced in the S-17 scenario, with the black line and the
    `$F_L(17)$' result corresponding to a fit to all points shown.}
    \label{F_Fits}
\end{figure}

Results with the different numbers of beam configurations
are compared at the level of the Rosenbluth fits for a
typical example
bin in $x$ and $Q^2$ in Figure~\ref{F_Fits} in the conservative scenario.
The larger number of beam energy configurations leads both to an increase
in the number of data points available for the fits and to an extended
lever-arm for the fit. For the example bin shown, the uncertainty on 
$F_L$ decreases by a factor of 
approximately three between the S-5 and S-9 scenarios
and by a further factor of around two when extending to the full S-17
case. 

\begin{figure}[htb]
    \centering  
    \includegraphics[scale = 0.2]{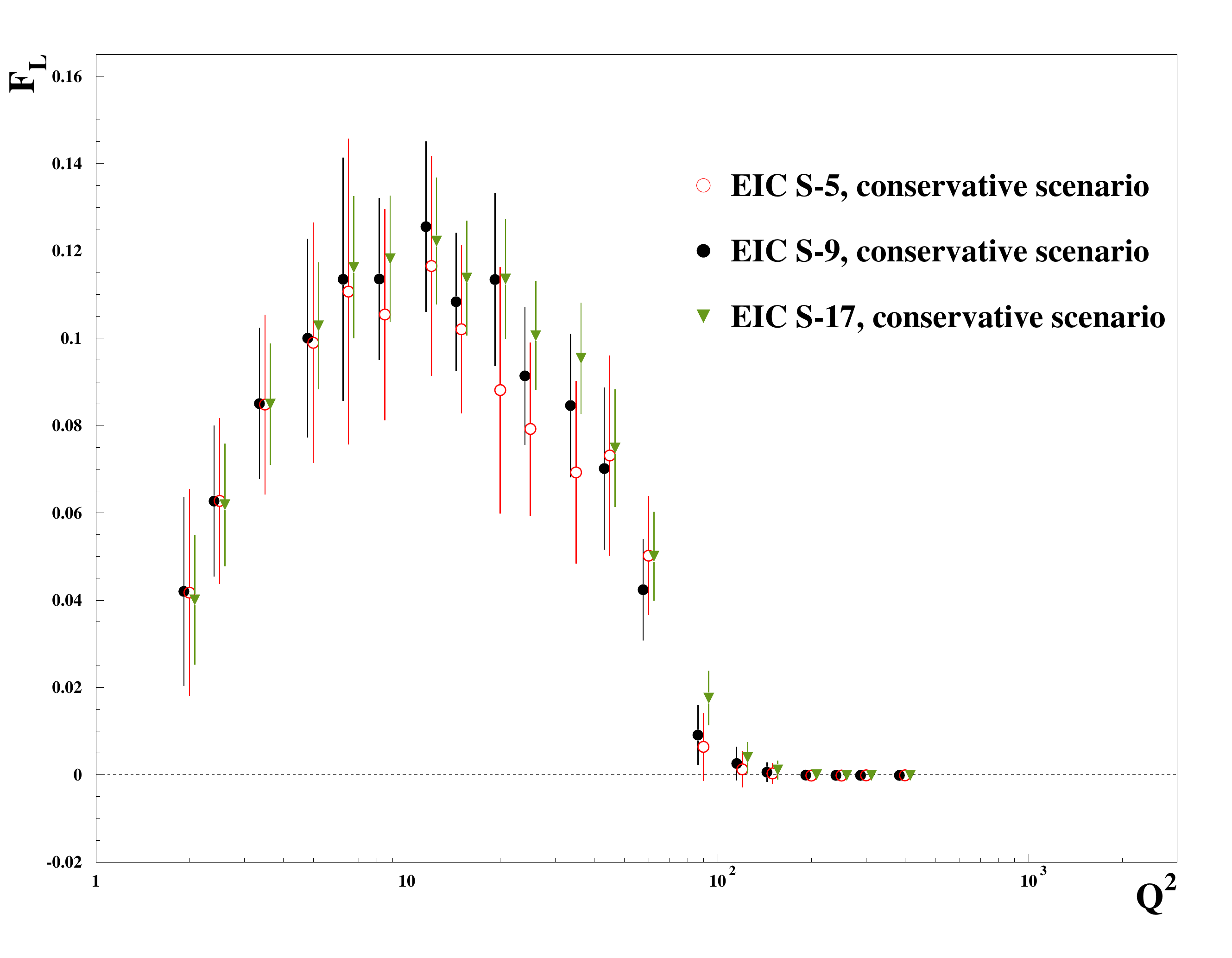}
    \includegraphics[scale = 0.2]{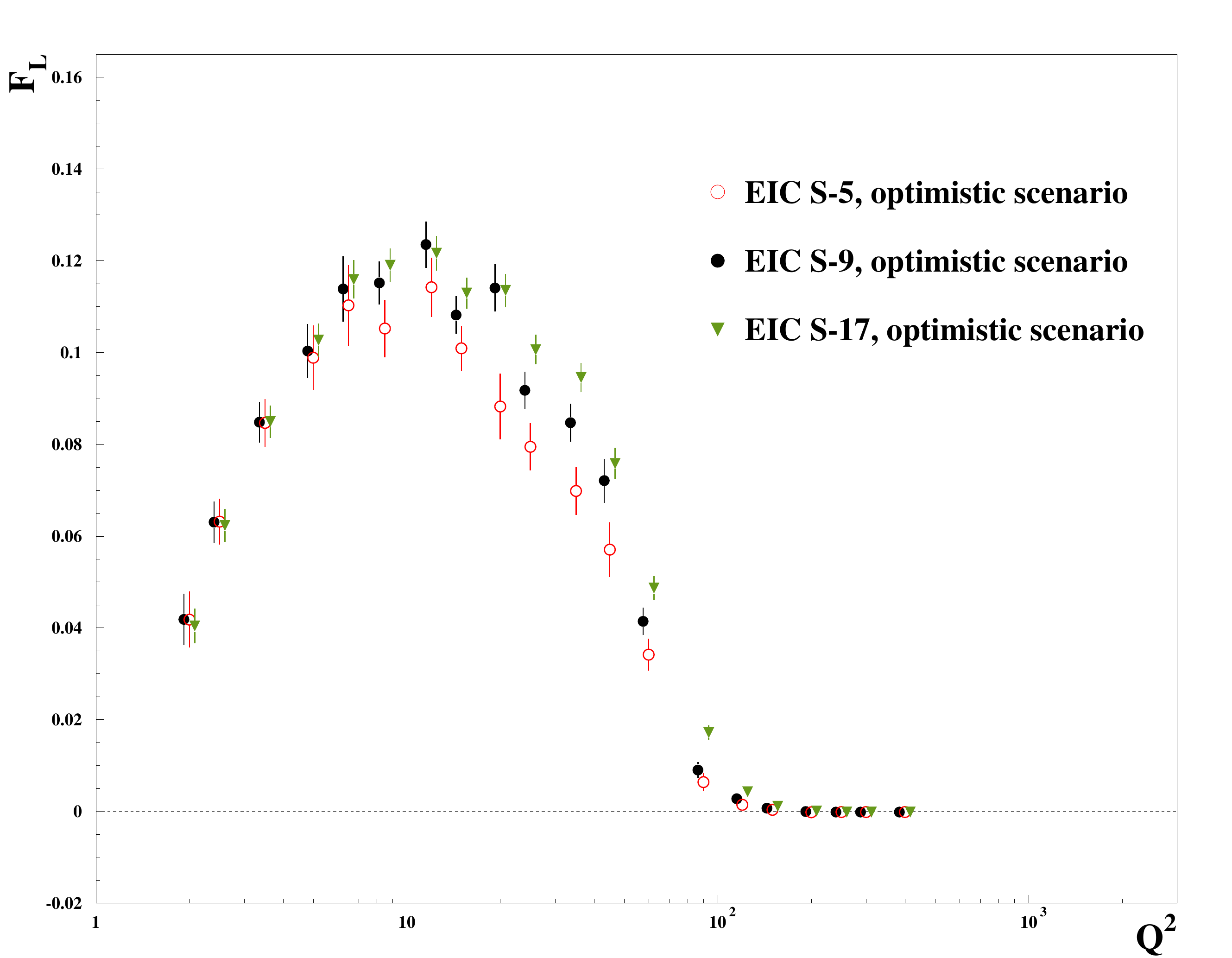}
    \caption{Comparison between simulated EIC $F_{L}(Q^2)$ results
    averaged over $x$ and shown as a function of $Q^2$ in the 
    S-5, S-9 and S-17 scenarios regarding beam energy combinations. The
    conservative assumption on systematic precision is shown on the 
    left, with the optimistic scenario on the right. Data points from
    the different scenarios are slightly shifted relative to one another 
    for visibility.}
    \label{Improvmax}
\end{figure}

As noted in Section~\ref{sec:replicas}, our method leads to
bin-to-bin fluctuations in the uncertainties on the extracted $F_L$
due to the differences between the 
number of accessible points and their $y^2 / Y_+$ ranges as $x$ and $Q^2$ are varied. 
This is reflected in the improvements in the uncertainties
in the two-dimensional projections as more $\surd s$ values are added,
with some bins unaffected and others showing substantial
improvements. 
An additional region of phase space at low 
$y$ also becomes accessible with reasonable precision. 
The impact of the additional beam energy configurations is illustrated
at the level of the $x$-averaged dependence of $F_L$ on $Q^2$ in 
Figure~\ref{Improvmax}
for both the optimistic and the conservative scenarios. 
The bin-to-bin fluctuations in the uncertainties are smoothed
out to a large extent in the averaging procedure. Significant 
improvements are visible in the precision both when increasing the
number of configurations from the S-5 to the S-9 set and also from
S-9 to S-17. This remains the case for both the optimistic and the
conservative scenarios, though the introduction of more configurations
is not sufficient to compensate for the difference between the 
two rather different uncertainty scenarios. 
In these figures, the averaged $F_L$ values are different in the
different scenarios, since adding new points in the Rosenbluth fits
alters the selection of $F_L$ data points passing the uncertainty
criteria for inclusion in the figures and the uncertainties on each
individual point, leading to changes in the weighted average 
values of $x$. 

The simple studies performed in this section represent 
only a small fraction of the total range of possible variations
that could be studied, and further optimisation is therefore certainly possible. 
Nevertheless, it is clear that 
minimising the 
systematic uncertainties 
on the data is the single most important
ingredient in obtaining high precision on $F_L$, and that
a dedicated running period with lots of beam energy 
configurations and modest luminosities 
is a possible approach to obtaining a high quality result with a
relatively small amount of EIC beam time. 

\section{Summary and context}
\label{sec:conclu}

In this work, we have explored the potential of the Electron-Ion Collider (EIC) to measure the longitudinal structure function $F_{L}$ 
using simulated data.
Our studies indicate that the EIC will be able to measure 
$F_{L}(x, Q^2)$ 
with an unprecedented precision over a wide kinematic range. 
In scenarios with the most favourable assumptions, 
absolute uncertainties of less than 0.05 are achievable across
a wide kinematic range, corresponding to around 20\% where
$F_L$ is largest and offering competitive constraints on the 
proton gluon density. 
This represents a significant improvement compared with HERA,
which was sensitive to a region where $F_L$ is larger than at
EIC, but where the most precise results 
from double-differential 
cross sections in $x$ and $Q^2$ have uncertainties of around
of 30-40\%, and where around 30\% of all 
measurements have uncertainties larger than 100\%.
No attempt has been made to optimise the binning schemes
used here. Refinements, for example by adopting a $(y, Q^2)$
instead of an $(x, Q^2)$ grid to optimally populate the 
$y^2/Y_+$ range in the Rosenbluth decomposition fits
are expected to lead to 
improved precision and to the addition of further bins at 
low $Q^2$ and high $x$ that are kinematically accessible but 
excluded in the current analysis on the basis of having very
large uncertainties on $F_L$. 

By making variations in the analysis details, we have investigated the
impact of different qualities of measurement and 
choices of EIC running plan.
The most striking dependence is on the 
magnitude of the systematic uncertainties 
that are not correlated between different $\surd s$ values at the 
same $x$ and $Q^2$. The rather aggressive $1\%$ `optimistic'
assumption  
for such uncertainties that was studied here 
may not to be ultimately achievable, but it is reasonable to 
expect the final EIC situation to be much closer to 1\% than to the 
`conservative' 3.9\% scenario that was also investigated.  
By comparison with the systematics, statistical uncertainties
play a much more minor role; a reflection of the fact that $F_L$ is
obtained from inclusive measurements and the large expected sample
sizes at the EIC. If the initial assumption of a year's running at each
beam energy is drastically reduced to an example 
$1 \ {\rm fb^{-1}}$ for each of the five beam energy configurations, the
expected precision on $F_L$ is not substantially deteriorated. 
On the other hand, a significant advantage can be gained by adding
more beam energy configurations, which leads to more data points 
in the
Rosebluth decomposition fits spanning a wider range in the relevant
$y^2 / Y_+$ variable. The precision improves progressively
when studying
cases in which either nine or 17 configurations
are included in place of the default five. 
Running scenarios with relatively low luminosities collected at
relatively large numbers of different
$\surd s$ values seem to be the most efficient in terms of precision on
$F_L$. 

\begin{figure}[htb]
    \centering  
    \includegraphics[scale = 0.4]{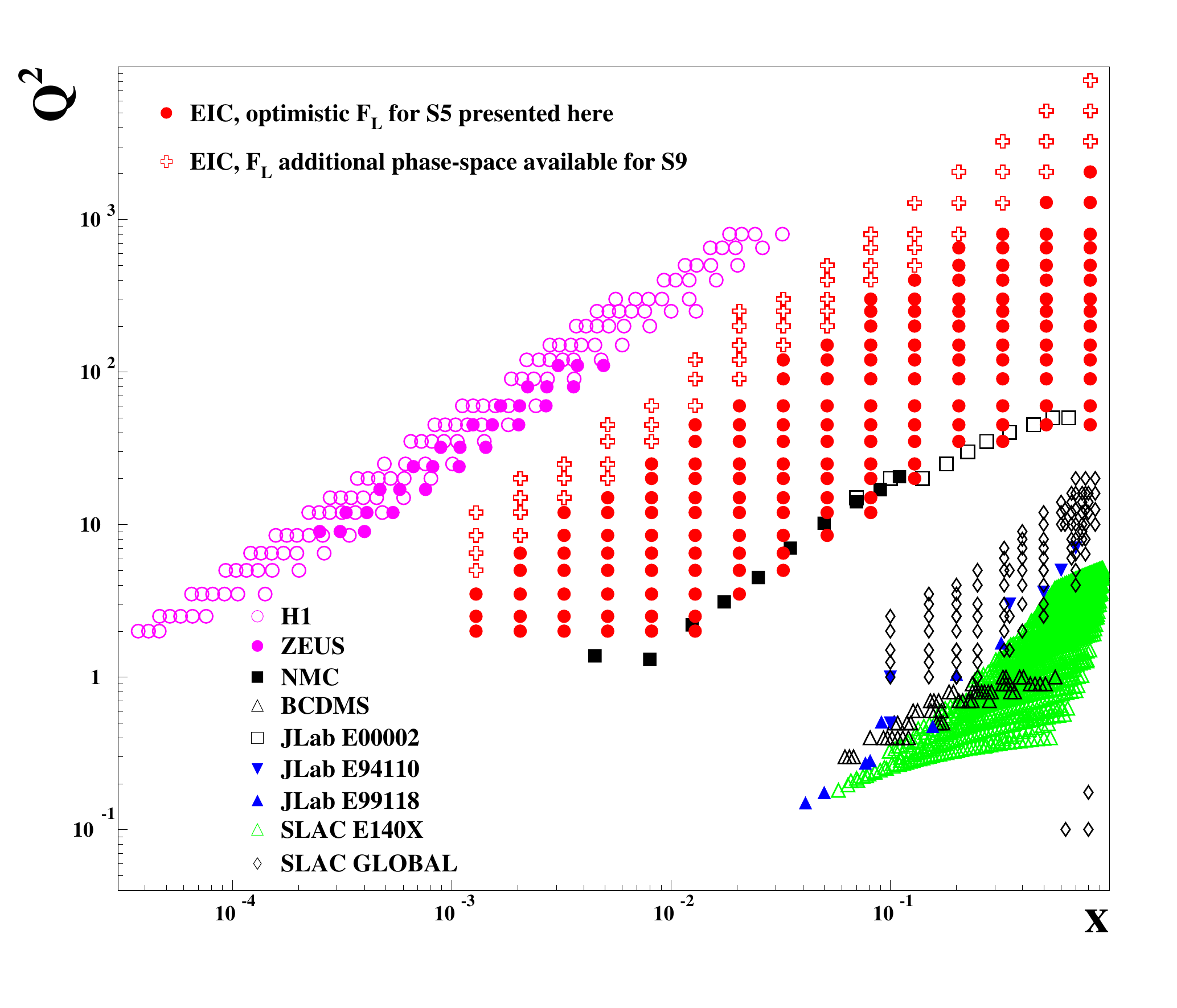}
    \caption{Kinematic coverage of world data for the proton longitudinal structure function, derived from figure 1 in
    \cite{PhysRevC.97.045204}, but with simulated EIC data added.
    The existing data shown are from H1 and ZEUS~\cite{H1:2013ktq,ZEUS:2014thn}, NMC~\cite{ARNEODO19973}, BCDMS~\cite{BENVENUTI1989485}, JLab~\cite{PhysRevC.97.045204,Tvaskis:2006tv,JeffersonLabHallCE94-110:2004nsn} and SLAC~\cite{E140X:1995ims,Whitlow:1991uw}. 
    The EIC pseudo-data are shown in the S-5 and S-9 scenarios, 
    which correspond to five and nine different beam energy configurations,
    respectively.}
     \label{Kplane-full}
\end{figure}

The kinematic region that is accessible with EIC data is 
complementary to that of HERA, covering a region of larger $x$ 
and also 
extending to higher $Q^2$.
The coverage of the 
EIC data is placed in the wider context of world $F_L$ data, 
also including fixed target measurements, in 
Figure~\ref{Kplane-full}. 
The EIC data close the large current gap
between the fixed target and the
HERA data. This is already the case in the 
scenario with five $\surd s$ values (S-5), but the extension
to nine
configurations (S-9), yields additional data points at large $Q^2$,
leading to a rather complete coverage of the kinematic
plane overall for $Q^2 > 1 \ {\rm GeV^2}$, extending as low as 
$x \sim 10^{-4}$ and approaching the situation for $F_2$. 
Altogether, the EIC has the potential to
transform our knowledge of $F_L$ across a wide kinematic
range, potentially
leading to a substantial impact on the precision on the 
proton PDFs and constraining phenomenological models in new ways.

\section*{Acknowledgements}

Much of the analysis
framework used here has evolved from collaboration with Nestor Armesto,
Anna Sta\'sto and Wojtek S\l{}omi\'nski, to whom we are grateful. 
We would also like to thank Oliver Schulz for his help with the BAT package. Javier Jim\'enez acknowledges the DESY Summer Student Program for significantly contributing to the development of this work.  

\bibliography{bib}

\end{document}